\documentclass[singlecolumn]{aastex63}
\usepackage{rotating}
\usepackage{graphicx}
\usepackage{amssymb}
\usepackage{amsmath}
\usepackage{hyperref}


\def\gtrsim{\mathrel{\hbox{\rlap{\hbox{\lower4pt\hbox{$\sim$}}}\hbox{$>$}}}}
\def\lesssim{\mathrel{\hbox{\rlap{\hbox{\lower4pt\hbox{$\sim$}}}\hbox{$<$}}}}
\def\farcs{\hbox{$.\!\!^{\prime\prime}$}}

\begin{document}

\title{
First ejection from the PSR B1259-63/LS 2883 high mass gamma-ray binary detected during the 2021-2024 binary cycle}
\author{Jeremy Hare}
\affiliation{NASA Goddard Space Flight Center, Greenbelt,  MD 20771, USA}
\affiliation{Center for Research and Exploration in Space Science and Technology, NASA/GSFC, Greenbelt, Maryland 20771, USA}
\affiliation{The Catholic University of America, 620 Michigan Ave., N.E. Washington, DC 20064, USA}
\author{George G.\ Pavlov}
\affiliation{Department of Astronomy \& Astrophysics, Pennsylvania State University, 525 Davey Lab, University Park, PA 16802, USA}
\author{Oleg Kargaltsev}
\affiliation{Department of Physics, The George Washington University, 725 21st St. NW, Washington, DC 20052}
\affiliation{The George Washington Astronomy, Physics, and Statistics Institute of Sciences (APSIS)}
\author{Gordon P.\ Garmire}
\affiliation{Huntingdon Institute for X-ray Astronomy, LLC, 10677 Franks Road, Huntingdon, PA 16652, USA}
\email{jeremy.hare@nasa.gov}

\begin{abstract}
LS 2883/PSR B1259-63 is a high mass, eccentric gamma-ray binary that has previously been observed to eject X-ray emitting material. We report the results of recent Chandra observations 
near binary apastron in which a new X-ray emitting clump of matter was detected.
The clump has a high projected velocity of $v_{\perp}\approx 0.07c$ and hard X-ray spectrum, which fits an absorbed power-law model with $\Gamma=1.1\pm0.3$.  Although clumps with similar velocities and spectra were detected in some of the previous binary cycles, no resolved clumps were seen near apastron in the preceding cycle of 2017-2021. 
\end{abstract}

\section{Introduction}
LS 2883/PSR B1259-63 (B1259 hereafter) is a high mass gamma-ray binary consisting of a 15-30 $M_{\odot}$ Be star and a neutron star  \citep{1992ApJ...387L..37J,2011ApJ...732L..11N,2018MNRAS.479.4849M}. The binary has a large eccentricity ($e=0.87$), an orbital period of about 1236.7 days (or 3.4 years), and is located at a distance of 2.6 kpc  \citep{2018MNRAS.479.4849M}. The 48 ms pulsar in this system has an energy loss rate $\dot{E}\approx8\times10^{35}$ erg s$^{-1}$ and a characteristic age $\tau = 330$ kyr. The massive Be star has a stellar wind, as well as an equatorial decretion disk formed due to its rapid rotation. It is thought that this decretion disk is inclined to the orbit of the pulsar, such that the pulsar passes through this disk twice near periastron passage \citep{1999ApJ...514L..39B}.

In previous binary cycles, variable extended X-ray emission 
was detected with Chandra 
at $\lesssim 10''$ south-southwest of the binary \citep{2011ApJ...730....2P,2014ApJ...784..124K,2015ApJ...806..192P,2019RLSFN..30S.125P,2019ApJ...882...74H}. Interestingly, the sources of this extended emission, nicknamed ``clumps'', were found to travel away from the binary at projected velocities of $\sim0.1c$, and they even  showed possible evidence of being accelerated \citep{2014ApJ...784..124K,2015ApJ...806..192P,2019ApJ...882...74H}. It has been hypothesized that these clumps may be material from the stellar decretion disk ejected by the pulsar's passage through the disk. This material then gets entrained into a channel in the stellar wind carved out by the pulsar wind (PW). Due to the high eccentricity of the binary, and the pulsar spending a majority of its orbit near apastron, the 
    PW is predominantly confined and directed towards the orbit's apastron, assuming that the Be star wind is dynamically dominant \citep{2013A&ARv..21...64D}. Once the debris from the disk enters the PW, it is then accelerated to the high observed velocities. In this scenario, the X-ray emission comes from relativistic electrons accelerated by the shock between the PW and the slower moving disk debris (see \citealt{2015ApJ...806..192P,2016MNRAS.456L..64B,2019ApJ...882...74H} for additional details).

\section{New Observational Results}

B1259's most recent periastron passage occurred on 2021 February 9.
 To search for new clumps being launched from the binary, we observed B1259 
four times with the Chandra ACIS-I detector near apastron for a total of $\approx47$ ks. 
The observations were carried out on 2022 November 18-19 
(ObsIDs 26011, 27557, 27558, 27559), i.e., 647-648 days after periastron passage. The observational setup was the same as that used in our previous observing campaigns \citep[see][for more details]{2015ApJ...806..192P,2019ApJ...882...74H}. 
To correct the relative astrometry, we aligned and stacked the images 
 using the position of the binary for alignment, with obsID 26011 as the reference image.

Extended X-ray emission is clearly seen to the south of the binary (see Figure 1, left panel). The extended emission is located about 3$''$ from the position of the binary and is about $3''\times0.8''$ in size although it does not appear to be entirely separated from the binary. Assuming the material responsible for the X-ray emission was launched at periastron and has been moving with a constant speed, its projected velocity is $v_{\perp}=0.07c$. 
This is similar to the mean velocity of the clumps launched during the 2010 and 2014 periastron passages\citep{2015ApJ...806..192P,2019ApJ...882...74H}. Interestingly, extended emission was hardly detectable in the 2017-2021 binary cycle (see an example in the right panel of Figure 1), so the velocity of the alleged clump could not be measured (Hare et al., in preparation).

We extracted the spectrum of the clump from each observation, using an elliptical extraction region with semi-major and minor axes $r_{\rm major}\approx2\farcs4$ and $r_{\rm minor}\approx0\farcs83$, respectively, and fit them simultaneously in the 0.5-8 keV energy band. In total, the clump had 57 net counts. We fit the spectrum of the point source with an absorbed power-law model to measure the absorbing column density and found $N_{\rm H}=(0.24\pm0.10)\times10^{22}$ cm$^{-2}$. We then froze the absorbing column density to this value and fit the clump spectra with an absorbed power-law model using C-statistics due to the small number of counts \citep{1979ApJ...228..939C}. The best-fit model (C-stat of 61.2 for 57 degrees of freedom) has a photon index $\Gamma=1.1\pm0.3$, and unabsorbed 0.5-8 keV flux $F_{\rm X}=(5\pm1)\times10^{-14}$ erg cm$^{-2}$ s$^{-1}$, which is consistent within errors 
with the spectra 
of previously detected clumps.

Further X-ray observations are needed to measure the extended emission's flux and morphological evolution and understand the mechanisms of clump formation, ejection and emission. Additionally, sensitive high angular resolution radio observations would help to constrain the
multi-wavelength spectrum of the clump.

\acknowledgements
This work is supported by Chandra ACIS Team contract SV4-74018 (G.\ P.\ Garmire \& L.\ Townsley, Principal Investigators), issued by the Chandra X-ray Observatory Center, operated by the Smithsonian Astrophysical Observatory for and on behalf of NASA under contract NAS8-03060.

\begin{figure*}
\includegraphics[trim={0 0 0 0},width=18.0cm]{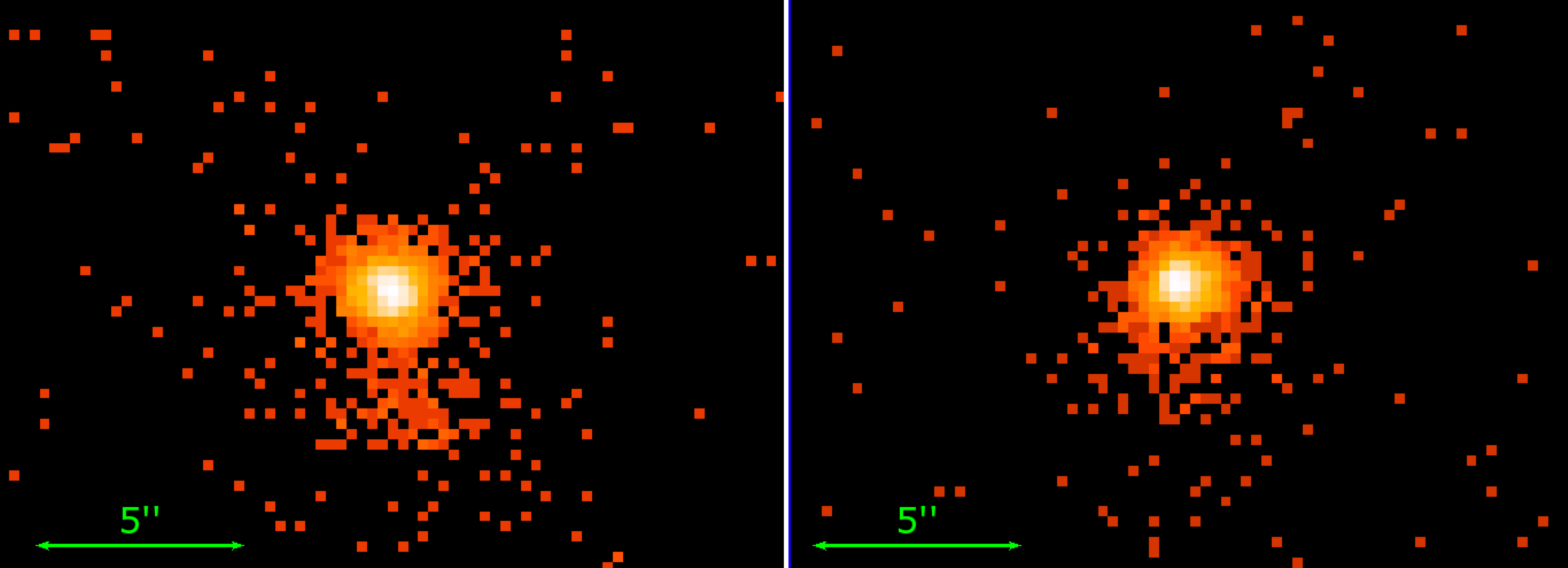}
\caption{Chandra ACIS-I images of B1259 with pixel size of 0.246$''$ 648 days after he 2021 periastron passage (left) and 610 days after the 2017 persiastron passage (right).}
{
\label{obs_fig}
}
\end{figure*}


\begin{thebibliography}{}

\bibitem[Ball et al.(1999)]{1999ApJ...514L..39B} Ball, L., Melatos, A., Johnston, S., et al.\ 1999, \apjl, 514, L39. doi:10.1086/311928

\bibitem[Barkov \& Bosch-Ramon(2016)]{2016MNRAS.456L..64B} Barkov, M.~V. \& Bosch-Ramon, V.\ 2016, \mnras, 456, L64. doi:10.1093/mnrasl/slv171




\bibitem[Cash(1979)]{1979ApJ...228..939C} Cash, W.\ 1979, \apj, 228, 939. doi:10.1086/156922

\bibitem[Dubus(2013)]{2013A&ARv..21...64D} Dubus, G.\ 2013, \aapr, 21, 64. doi:10.1007/s00159-013-0064-5




\bibitem[Hare et al.(2019)]{2019ApJ...882...74H} Hare, J., Kargaltsev, O., Pavlov, G., et al.\ 2019, \apj, 882, 74. doi:10.3847/1538-4357/ab3648



\bibitem[Johnston et al.(1992)]{1992ApJ...387L..37J} Johnston, S., Manchester, R.~N., Lyne, A.~G., et al.\ 1992, \apjl, 387, L37. doi:10.1086/186300




\bibitem[Kargaltsev et al.(2014)]{2014ApJ...784..124K} Kargaltsev, O., Pavlov, G.~G., Durant, M., et al.\ 2014, \apj, 784, 124. doi:10.1088/0004-637X/784/2/124


\bibitem[Miller-Jones et al.(2018)]{2018MNRAS.479.4849M} Miller-Jones, J.~C.~A., Deller, A.~T., Shannon, R.~M., et al.\ 2018, \mnras, 479, 4849. doi:10.1093/mnras/sty1775


\bibitem[Negueruela et al.(2011)]{2011ApJ...732L..11N} Negueruela, I., Rib{\'o}, M., Herrero, A., et al.\ 2011, \apjl, 732, L11. doi:10.1088/2041-8205/732/1/L11

\bibitem[Pavlov et al.(2011)]{2011ApJ...730....2P} Pavlov, G.~G., Chang, C., \& Kargaltsev, O.\ 2011, \apj, 730, 2. doi:10.1088/0004-637X/730/1/2

\bibitem[Pavlov et al.(2015)]{2015ApJ...806..192P} Pavlov, G.~G., Hare, J., Kargaltsev, O., et al.\ 2015, \apj, 806, 192. doi:10.1088/0004-637X/806/2/192

\bibitem[Pavlov et al.(2019)]{2019RLSFN..30S.125P} Pavlov, G.~G., Hare, J., Kargaltesev, O. 2019, Rendiconti Lincei. Scienze Fisiche e Naturali. Vol.\ 30, Issue Suppl 1, p.125-129. doi:10.1007/s12210-019-00765-0








\end{thebibliography}
\end{document}